\documentstyle[aps,graphicx]{revtex}
\begin{document}
\twocolumn
%%\draft
%%\tighten
{\bf Comment on ``Phase Coexistence in Multifragmentation''}\\
In their letter Moretto et al. \cite{moretto96} propose the fragment charge
distribution in nuclear multifragmentation to give a signal for the coexistence
of nuclear liquid and vapor phase. To our opinion this signal is not usefull
and misleading as fluctuations of different origin spoil it.

Phase transitions in macro-physics are usually indicated by a peak in the
specific heat e.g. $c_p(T)$ or by an anomaly of the caloric equation of state
($C\!E\!S$) $T(E)$ e.g. at constant pressure or volume. 
In closed finite systems, as e.g. highly excited nuclei,
phase transitions are well indicated by the shape of the $C\!E\!S$
c.f.\cite{gross72,gross95}.
Inherent to phase transitions are large fluctuations at the transition
which do not allow a clear phase separation in space or any other observable in
small finite systems because of the nonvanishing coherence length of the phase
fluctuations c.f. \cite{janke95,gross150} and which differ
at const.E and at const.T.
E.g. even though the backbending of the
$C\!E\!S$ is clearly seen for a 10-state Potts model at a lattice size of
$100*100$ and the area under the oscillation of $T^{-1}(E)$ is close to the
asymptotic value of the surface entropy no phase separation can be seen in the
configurations. Therefore, the interpretation by ref.\cite{moretto96}
is too naive and suffers from several further difficulties:

Equations like formulas (1-3) of ref.\cite{moretto96} notice charge
conservation only via the mean value but leave its fluctuation unrestricted. These
fluctuations are usually substantial especially near to phase transitions.
Moreover, in nuclear fragmentation one has to take care of the
indistinguishability of identical fragments and the partition problem is not
the Euler problem as is suggested in \cite{phair95}.  The correct formula for
the quantum partition of an integer $Z_0$ is given in
\cite{gross110}.

The experimental method used in ref.\cite{phair95,moretto96} is of course not
ideally suited to look for a phase transition in equilibrized nuclear systems.
First of all this system is generated in a collision of two sizeable nuclei.
The transverse energy $E_t$ does not give the total excitation energy
$\varepsilon^*$ of the system nor is it neccessarily proportional to it. In
fact the width in $\varepsilon^*$ at low $E_t$ can easily be of the order of
%$5$
a few MeV/nucleon\cite{moretto94a}. I.e. a fixed value of $E_t$ allows for
considerable fluctuations of $\varepsilon^*$.  It is therefore neccessary to
investigate the signal of ref.\cite{moretto96} in a situation where we
definitely {\em have an equilibrized} nuclear system with a {\em sharply
defined excitation energy}.

Lacking experimental data of this kind we investigated a model system from
which we know by experience that it reproduces nuclear multifragmentation and
which shows a nuclear phase-transition of first order towards fragmentation:
The Berlin microcanonical Metropolis Monte Carlo $M\!M\!M\!C$-model
\cite{gross95}.  Here as also in other versions of statistical
multifragmentation models like the Copenhagen model \cite{bondorf95} the  phase
transition towards fragmentation is clearly seen as anomaly in the $C\!E\!S$.

In fig.1 we show the resultant parameter $cZ$ or better the quantity
$cZ=ln\{P_n(Z)/P_{n+1}(Z)\}$ averaged over the IMF-multiplicities $n$ to
get better statistics vs.$Z$. Here $P_n(Z)$ is the probability to find one
fragment with charge $Z$ in events with n IMF's.  The two panels at
$\varepsilon^*=5$ and 
$=6$ MeV/nucleon resemble the findings of Moretto. At lower excitation
in contrast to ref.\cite{moretto96} $cZ$ is forced to rise with $Z$ as the
emission of a second fragment with larger $Z$ is prohibited due to limited
energy resources. 
\begin{figure}
\vspace{-5mm}
\includegraphics*[bb = 140 17 476 360, angle=-180, width=5.5cm,  %<-- altern. method
clip=true]{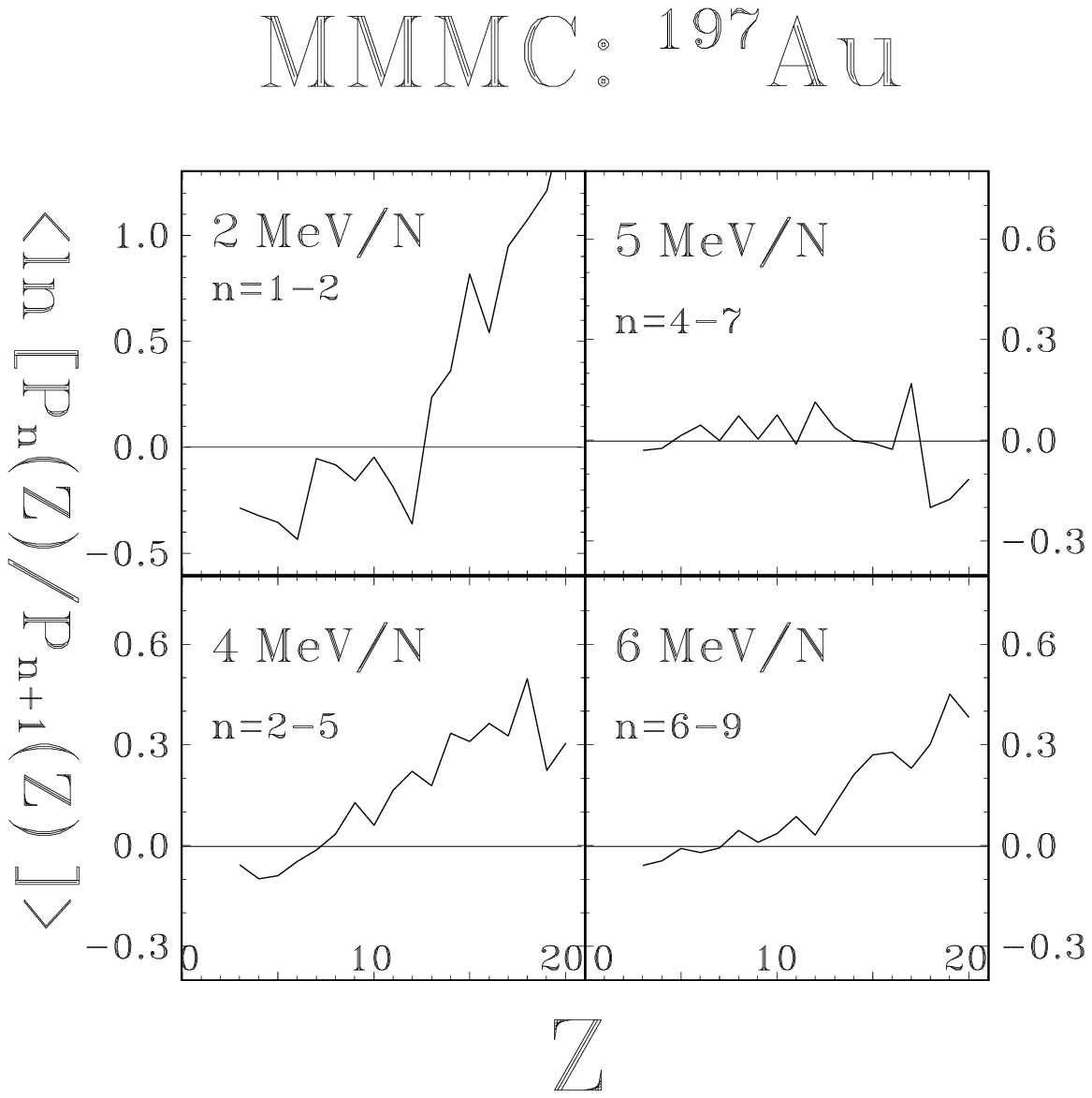}                                         %<--
% the four numbers are x-lower-left y-lower-left x-upper-right y-upper-right
%    in postscript units as read from ghostview
\end{figure}
We guess that at low {\em transverse} energy the experimental data of
ref.\cite{moretto96} are overshadowed by deep inelastic collisions
where some of the small fragments are likely from projectile
break-up which as such have small transverse energies. Consequently, low
total transverse energies do not really characterize the limitation to low
excitation energies as indicated by the large width in $\varepsilon^*(E_t)$
\cite{moretto94a}. This is probably the reason for the vanishing quantity $c$
found in ref.\cite{moretto96} at low transverse energies.

Conclusion: From all experience of microcanonical first order
phase transitions in small systems one knows that it is normally rather
difficult to see a clear phase separation even though the caloric equation of
state gives an unambiguous signal, phase fluctuations are usually too large.
Within the arguments of ref.\cite{phair95,moretto96} there are at least {\em
two} important conservation laws to be observed by the reaction: Conservation
of charge {\em and energy}. The latter forces the ``chemical potential'' $c$ to
rise again at low excitation energy.  The observation of an anomaly in the
caloric equation of state \cite{pochodzalla95} is still the best signal for a
phase transition as was predicted in \cite{gross72,bondorf95,gross95}. Since
long this is one of the classical signals for phase-transitions.\\~\\
D.H.E.~Gross and A.S.~Botvina 
Hahn-Meitner-Institut
Berlin, Bereich Theoretische Physik,14109 Berlin, Germany\\
%\bibliographystyle{unsrt}%{alpha}%{plain} %{unsrt}
%\bibliography{c:/bibliogr/gross,c:/bibliogr/othbibam,c:/bibliogr/othbibnz} 
%\end{document}
\vspace{-2.4cm}
%\rule{0mm}{-2cm}

\end{document}